\documentclass[aps,prl,twocolumn,floatfix,superscriptaddress]{revtex4}

\usepackage{epsfig}
\usepackage{amsmath}

\begin{document}
\preprint{DUKE-TH-02-224}

\title{Net baryon density in $Au+Au$ collisions at the Relativistic 
       Heavy Ion Collider}
\author{Steffen A.~Bass}
\affiliation{Department of Physics, Duke University, 
             Durham, North Carolina 27708-0305}
\affiliation{RIKEN BNL Research Center, Brookhaven National Laboratory, 
             Upton, New York 11973, USA}
\author{Berndt M\"uller}
\affiliation{Department of Physics, Duke University, 
             Durham, North Carolina 27708-0305} 
\author{Dinesh K.~Srivastava}
\affiliation{Variable Energy Cyclotron 
Centre, 1/AF Bidhan Nagar, Kolkata 700 064, India}            
\affiliation{Physics Department, McGill University,
             3600 University Street, Montreal, H3A 2T8, Canada} 
\date{\today}

\begin{abstract}
We calculate the net baryon rapidity distribution in Au+Au collisions
at the Relativistic Heavy Ion Collider (RHIC) in the framework of the
Parton Cascade Model (PCM). Parton rescattering and fragmentation leads
to a substantial increase in the net baryon density at mid-rapidity
over the density produced by initial primary parton-parton scatterings.
The PCM is able to describe the measured net baryon density at RHIC.
\end{abstract}

\maketitle
First experiments at the Relativistic Heavy Ion Collider (RHIC)
have shown that the matter created in the central rapidity
region contains a significant excess of baryons over antibaryons.
While a slight baryon excess was not unexpected, the magnitude of 
the net baryon multiplicity density 
$dN_{{\rm B}-\overline{\rm B}}/dy \approx 19 \pm 2$ \cite{bbar_exp,itzhak}
at $\sqrt{s}_{NN}=130$ GeV and $\approx 14 \pm 4$ \cite{brahms} 
at $\sqrt{s}_{NN}=200$ GeV 
is higher than what many theoretical models had predicted 
\cite{LC4RHIC}. In particular models in which
the deposition of energy
at midrapidity is driven by quasiclassical glue fields
\cite{ColorGlass} or fragmenting color flux tubes \cite{Schwinger}, 
which produce quarks and anti-quarks in equal abundance, 
underpredicted the data.  On the other hand, 
models invoking baryon junctions \cite{Kharzeev,HijingB}
for the transport of baryon number from 
the beam rapidity into the central region $y\approx 0$ have done
remarkably well. The baryon junction
mechanism was originally proposed as a means to understand baryon
number annihilation and stopping in elementary $p+p$ and $p+\bar p$ 
reactions \cite{junction_orig}.
In this letter we shall address the question whether only baryon 
junctions provide a mechanism capable of explaining the RHIC data, 
or whether they can also be understood in the framework
of a more conventional picture, based on parton distributions and 
pQCD driven multiple interactions.

Two effects based on established physics can contribute 
to the baryon excess at midrapidity. First, the measured 
parton distribution functions in the nucleon are well
known to exhibit a substantial asymmetry between quark
and antiquark distributions at moderately small values
of Bjorken-$x$ ($x\approx 0.01$) relevant for RHIC energies 
( Fig.~\ref{fig1}, see below). Second, it is known
from experiments with $p+A$ collisions that multiple
scattering is quite effective in transporting baryons
to smaller rapidities \cite{Busza}. The parton cascade 
model (PCM) \cite{GM92,Geiger,VNIBMS} provides a natural
framework for exploring the consequences of these two
effects quantitatively.

Devised as a description of the early, pre-equilibrium 
phase of a nucleus-nucleus collision at relativistic energy, 
the PCM does not include a description of hadronization 
and the subsequent scattering among hadrons.
These late-stage processes, however, are not expected 
to alter the distribution of net baryon number with
respect to rapidity, because baryon diffusion in a
hadronic gas is slow \cite{Stephanov} and net baryon number 
is locally conserved. We therefore believe that these 
limitations of the PCM approach should not stand in the 
way of an adequate explanation of the net baryon distribution.
In this work, we present calculations of the net baryon 
multiplicity density distribution in Au+Au collisions at 
130 and 200 GeV center-of-mass energy per nucleon pair. 

The fundamental assumption underlying the PCM is that the state 
of the dense partonic system can be characterized by a set of 
one-body distribution functions $F_i(x^\mu,p^\alpha)$, where $i$
denotes the flavor index ($i = g,u,\bar{u},d,\bar{d},\ldots$)
and $x^\mu, p^\alpha$ are coordinates in the eight-dimensional
phase space. The partons are assumed to be on their mass shell,
except before the first scattering. In our numerical implementation, 
the {\sc GRV-HO} parametrization \cite{grv} is used, and the parton 
distribution functions are sampled at an initialization scale 
$Q_0^2$ to create a discrete set of particles. Partons generally 
propagate on-shell and along straight-line trajectories between 
interactions. Before their first collision, partons may have a 
space-like four-momentum, especially if they are assigned an 
``intrinsic'' transverse momentum.  

The time-evolution of the parton distribution is governed by a 
relativistic Boltzmann equation:
\begin{equation}
p^\mu \frac{\partial}{\partial x^\mu} F_i(x,\vec p) = {\cal C}_i[F]
\label{eq03}
\end{equation}
where the collision term ${\cal C}_i$ is a nonlinear functional 
of the phase-space distribution function. The calculations discussed
below include all lowest-order QCD scattering processes between 
massless quarks and gluons \cite{Cutler.78}.
A low momentum-transfer cut-off $p_T^{\text{min}}$ is needed to 
regularize the infrared divergence of the perturbative parton-parton 
cross sections.  Additionally, we include the branchings $q \to q g$, 
$q \to q\gamma$, $ g \to gg$ and $g \to q\overline{q}$ \cite{frag}.  
The soft and collinear singularities in the showers are avoided by 
terminating the branchings when the virtuality of the time-like 
partons drops below $\mu_0 = 1$ GeV.  Some of these aspects were 
originally discussed in~\cite{Geiger}.  The present work is based 
on our thoroughly revised, corrected, and extensively tested 
implementation of the parton cascade model, named {\sc VNI/BMS}~\cite{VNIBMS}.

\begin{figure}[tb]   
\centerline{\epsfig{file=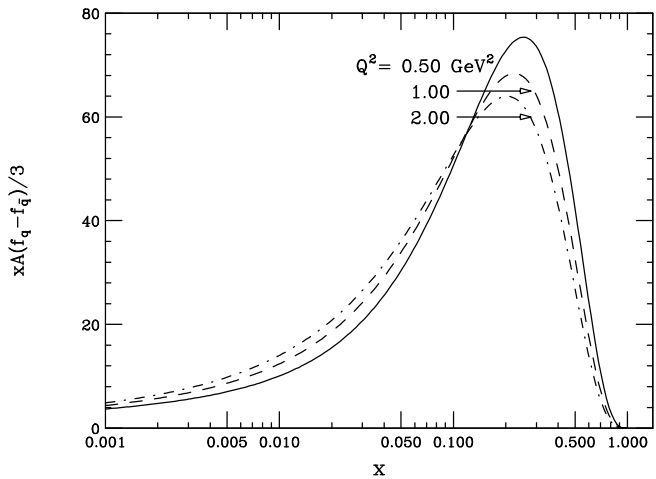,width=8.6cm}}
\caption{Net baryon content of the partonic distribution function
of gold-nucleus at the factorization scales ($Q_0^2$) of
0.50 GeV$^2$ (solid curve), 1.00 GeV$^2$ (dashed curve) and 2.00 GeV$^2$
(dot-dashed curve), for the {\sc GRV-HO}~\cite{grv} parametrization.
Parton shadowing is not included. }
\label{fig1}
\end{figure}

Figure~\ref{fig1} shows the net baryon-number distribution 
$x \,A \left[ f_{q}(x)-f_{\overline{q}}(x) \right ]/3$ with the 
Bjorken-$x$ variable. We have defined,
\begin{equation}
x f_q(x)=\sum_{i=u,d,s} x F_i(x)~,
\end{equation}
where $F_i(x)$ is the iso-spin averaged parton-distribution function 
for a nucleon in the gold nucleus, so that
\begin{equation}
F_i(x)=\frac{Z}{A}F_i^{P}(x)+\frac{(A-Z)}{A}F_i^{N}(x)~.
\end{equation}
where the superscripts $P$ and $N$ stand for protons and neutrons.
Thus the figure brings out the asymmetry between the quark and 
antiquark  distributions at moderately small values of Bjorken-$x$ 
($x\approx 0.01$) relevant for energy deposition at midrapidity at 
RHIC energies. The various lines denote different initialization 
scales for the parton distribution functions. This provides the 
basis for the evaluation of the dynamical evolution of the stopping 
through parton collisions, sometimes accompanied by gluon radiation 
and quark pair creation.

The partons in a fast moving nucleon or a nucleus are distributed 
both in the transverse and longitudinal directions. Due to the 
large longitudinal Lorentz contraction, the longitudinal 
momentum $p_z=xP$ is the most interesting variable.  Before or 
immediately after the first collision of a parton, one can relate 
$p_z$ to the rapidity variable $y = Y + \ln x + \ln(M/Q_s)$, 
where $Y$ is the rapidity of the fast moving nucleon, $M$ is the
nucleon mass, and $Q_s$ denotes the typical transverse momentum
scale. Depending on the picture of the initial state, $Q_s$ is 
either given by the average intrinsic virtuality, often called the
saturation scale \cite{sat-Q}, or by the typical transverse momentum
given to the parton in the first interaction which brings it onto
the mass shell. In any case, $|\ln(M/Q_s)| < 1/2$ for Au+Au 
collisions at RHIC.

If the partons in the two gold nuclei were to decohere completely 
upon passing through each other without any further interaction, 
the resulting net baryon rapidity distribution would be
approximately given by figure~\ref{fig1} -- predicting a contribution 
to the net baryon density of about 7 at mid-rapidity (with $Y=5.4$ 
for $\sqrt{s}_{NN} = $200 GeV) due to each nucleus. This is seen as 
follows. The baryon number in the nucleus $A$ is given by
\begin{equation}
A = \frac{1}{3} \, A \,\int \left[
    x\, f_q(x)- x\, f_{\overline{q}}(x)\right] \, d\ln x ~,
\end{equation}
so that the net baryon-distribution ``contained'' in the nucleus is
\begin{equation}
\left(\frac{dN}{dy}\right)_{B-\overline{B}}=\frac{1}{3}x \, A \,
\left[ f_q(x)-f_{\overline{q}}(x) \right]~.
\end{equation}
We shall see that the parton-interactions incorporated in the PCM 
remain consistent with this intuitive picture.

A crucial parameter of the PCM is the low momentum transfer cut-off 
$p_T^{\text{min}}$.  Under certain assumptions this parameter 
can be determined from experimental data for elementary 
hadron-hadron collisions \cite{sjostrand,naga,eskola1}.
In the environment of a heavy-ion collision, color screening 
will destroy the association of partons to particular hadrons,
since the density of free color charges is so high that the color
screening length becomes smaller than the typical hadronic scale.
In a previous publication \cite{VNIBMS}
we have established a consistency limit
for the allowed range of $p_T^{\text{min}}$ by calculating the screening 
mass $\mu_D$ in the produced parton matter as a function of the 
cut-off $p_T^{\text{min}}$ and demanding ($p_T^{\text{min}}\ge \mu_D$). 
The initialization scale and low momentum cut-off 
of the pQCD cross sections were chosen as 
$Q_0^2 = (p_T^{\text{min}})^2 = 0.50$~GeV$^2$ for $\sqrt{s}_{NN}=130$~GeV 
and 0.59~GeV$^2$ for $\sqrt{s}_{NN}=200$~GeV, respectively. 
The energy scaling of this parameter agrees with that determined 
by Eskola et al. \cite{EKRT} for the geometric minijet saturation 
model. 

\begin{figure}[tb]   
\centerline{\epsfig{file=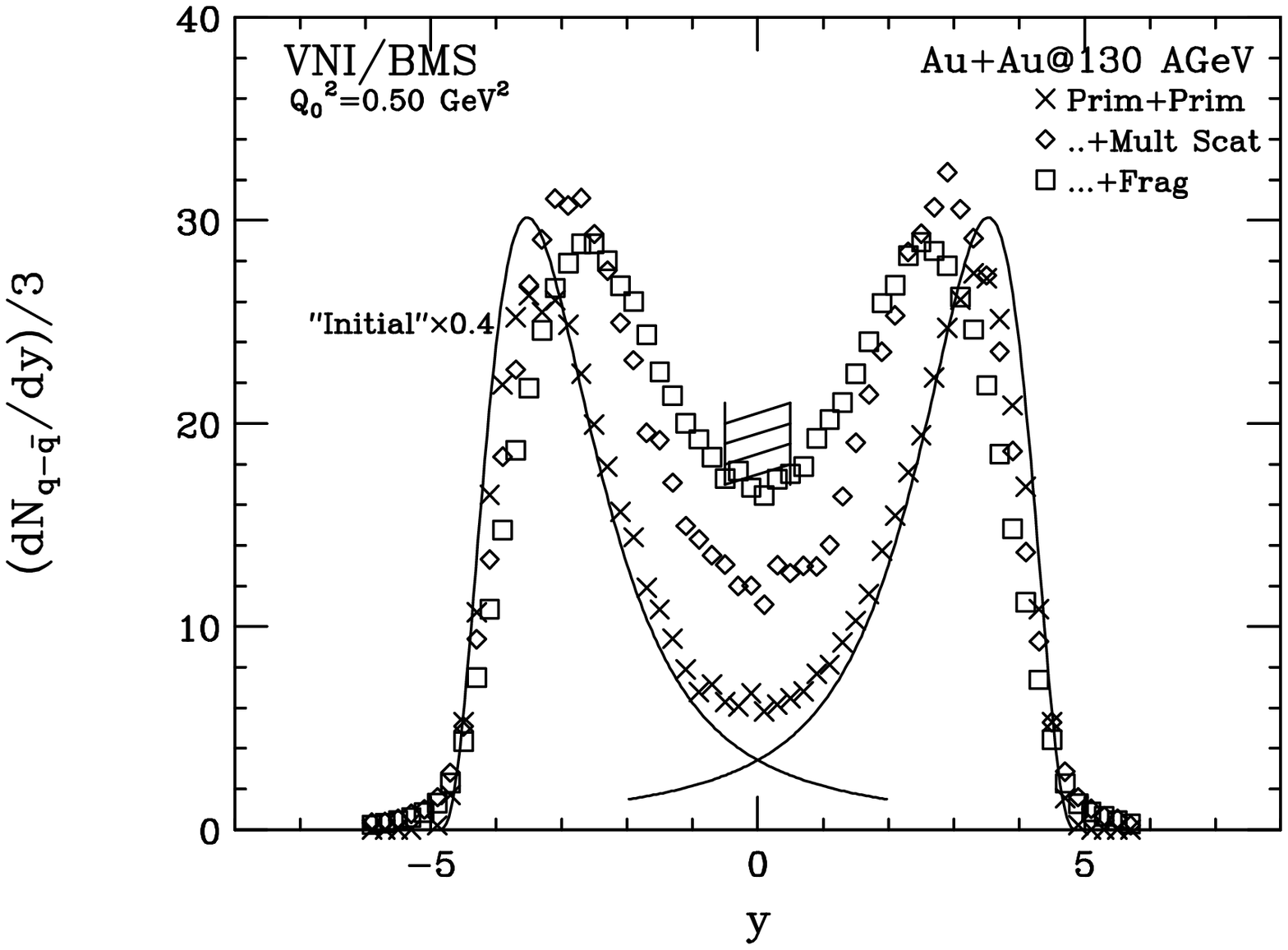,width=8.6cm}}
\centerline{\epsfig{file=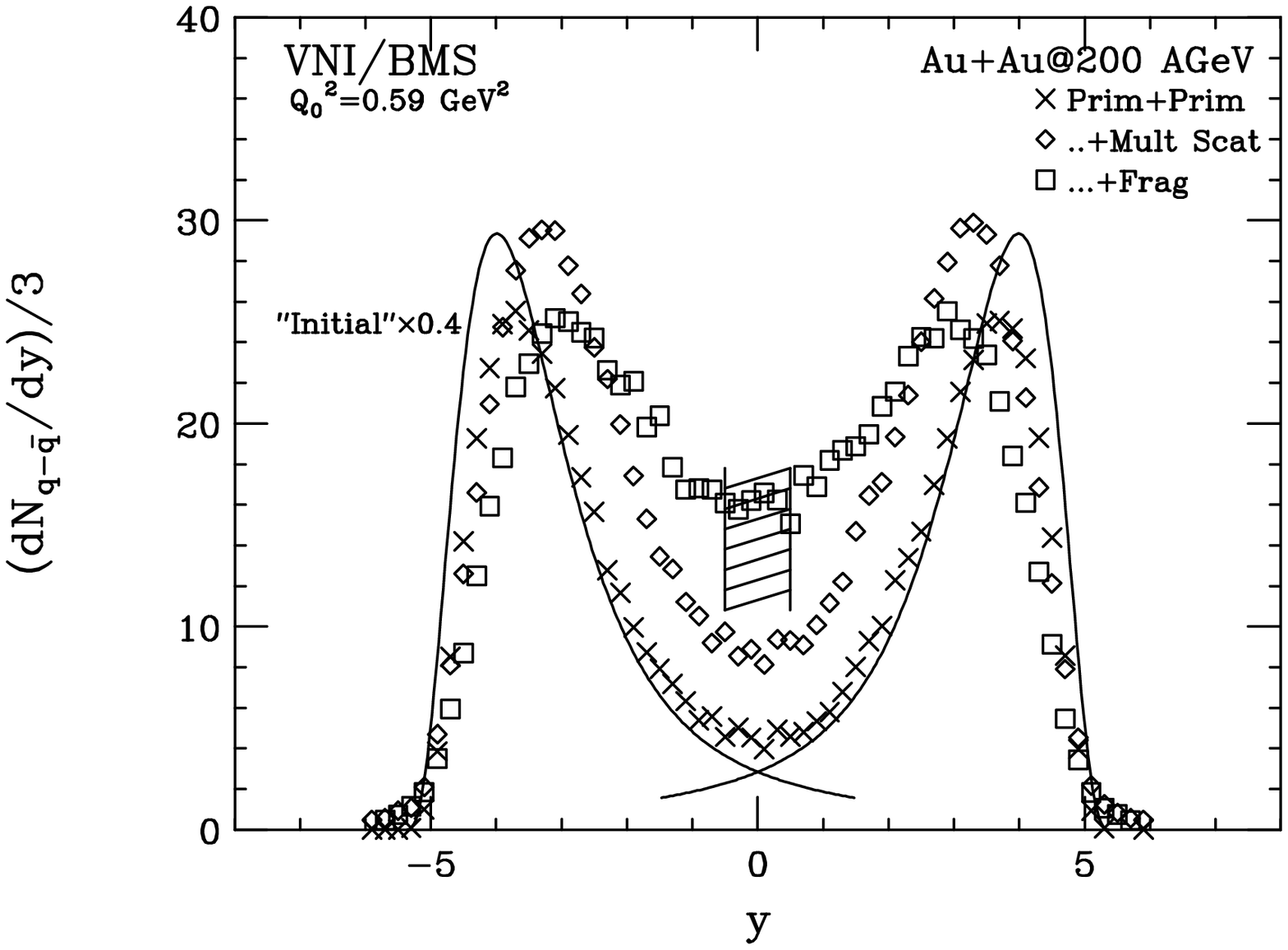,width=8.6cm}}
\caption{net baryon rapidity distributions for Au+Au reactions at
$\sqrt{s}_{NN}=130$~GeV (top) and $\sqrt{s}_{NN}=200$~GeV bottom.
 Crosses denote
a calculation in which the PCM has been restricted to primary parton
scatterings, diamonds include parton rescattering and squares include
rescattering and parton fragmentation. The solid lines show the
net baryon content of the partonic distribution functions for gold 
nuclei, scaled by an average liberation factor of 0.4 . The band
at $y_{CM}$ shows the range of experimental estimates for
the net-baryon density by STAR, BRAHMS and PHENIX.}
\label{fig2}
\end{figure}

Figure~\ref{fig2} shows the PCM predictions for the net baryon 
rapidity distributions for $\sqrt{s}_{NN}=130$~GeV (upper panel) 
and 200~GeV (lower panel), respectively.
The initial projectile and target rapidities are  
 $\pm 4.9$ for $\sqrt{s}_{NN}=130$~GeV and 
$\pm 5.4$ for $\sqrt{s}_{NN}=200$~GeV. 
The calculations are done without assigning any intrinsic $k_T$ 
to the partons -- initially all partons move with the velocity 
$\beta=\pm P_z^{\rm A}/E_A$ of the nucleus in the center-of-mass 
frame, where $P_z$ is the momentum and $E_A$ is the energy of the 
nucleus.

Crosses in Fig.~\ref{fig2} denote a calculation in which the 
PCM has been restricted to primary-primary parton scatterings, 
and therefore reflects a calculation in which each parton is 
allowed to scatter only once. Already one hard collision is
sufficient to deposit a net surplus of quarks into the mid-rapidity 
region, resulting in a net baryon density at $y_{cm}=0$ of 6.3 
for 130~GeV and 5.0 at 200~GeV. For comparison, the net baryon
number distribution for each colliding nucleus, scaled by a factor
0.4 from the distribution shown in Fig.~\ref{fig1}, is shown as
a solid line. The remarkable agreement demonstrates that the
net baryon number distribution produced by first parton-parton
collisions is predetermined by the initial parton structure of
the nuclei. The factor 0.4 is the average ``liberation factor'' $c$
for partons in the PCM for the selected parameters. 
This factor is 
consistent with predictions
by some gluon saturation models 
\cite{KV01}.

The diamonds in Fig.~\ref{fig2} represent a calculation with 
full parton-parton rescattering. Allowing for multiple parton 
collisions increases the net baryon density at mid-rapidity roughly 
by a factor of two at 130~GeV and by 75\% at 200~GeV, filling up the 
dip around mid-rapidity. This trend continues when parton fragmentation
is included (squares): at 130~GeV fragmentation processes add another
50\% to the net baryon density at mid-rapidity, bringing it up to 
about 18, whereas at 200~GeV the net baryon density increases to 
near 14. The rapidity change of a quark in each subsequent collision 
after its liberation in the first hard scattering yields an average 
rapidity shift of roughly 0.65 units per collision.

The band around mid-rapidity in Fig.~\ref{fig2} denotes the range of 
experimental estimates for the net-baryon density at mid-rapidity for
130~GeV and 200~GeV, respectively \cite{bbar_exp,brahms,itzhak}.
These estimates depend on how the extrapolations
from ($p-\bar p$) and ($\Lambda - \bar\Lambda$) to ($B - \bar B$) have been
carried out, the extreme scenarios being the assumption of full
isospin equilibration for baryons, $(B - \bar B) = 2 \times ( p - \bar p)$,
vs. the full inclusion of the isospin asymmetry of the initial state,
$(B - \bar B) = A/Z \times ( p - \bar p)$. Note that our calculation
does not suffer from this uncertainty.

\begin{figure}[tb]   
\centerline{\epsfig{file=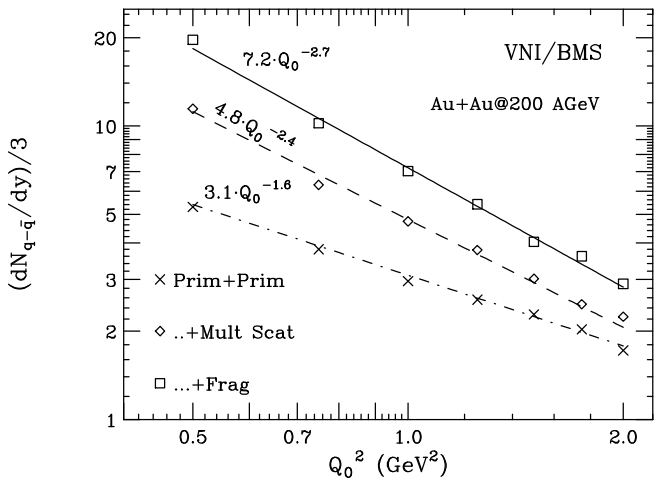,width=8.6cm}}
\caption{Initialization scale and cut-off dependence of the 
net baryon density at mid-rapidity for Au+Au collisions at
$\sqrt{s}_{NN}=200$~GeV.}
\label{fig3}
\end{figure}

Figure~\ref{fig3} shows the net baryon density at mid-rapidity as a 
function of the momentum cut-off $p_T^{\text{min}}$ for Au+Au collisions at 
$\sqrt{s}_{NN}=200$~GeV. Again, crosses denote the calculation restricted 
to primary-primary collisions, diamonds represent the full rescattering 
mode and squares include the effect of parton fragmentations. The 
observed power law dependence of the net baryon density as a function
of $Q_0$ stems from the properties of the pQCD cross sections in the
PCM. The absence of a saturation at small values of $Q_0$ indicates 
that not all valence quarks are ``liberated'' in the range of cut-off values
considered here. Indeed, we find that the liberation factor for quarks
in the nuclear parton distributions varies from about 0.7 for $x>0.1$
to about 0.2 for $x\approx 0.01$. Figure~\ref{fig3} can be used to 
rescale the PCM prediction for the net baryon rapidity distribution 
in Fig.~\ref{fig2} (bottom part) to other values of $Q_0$. For the
range of $Q_0$ values extracted from RHIC data the PCM is well able 
to describe the measured net baryon excess.

This success raises the obvious question: what does our model 
predict for the net baryon distribution at lower energies? 
We have reported~\cite{prc_contrast} recently that the
partonic cascades provide only a dilute medium at SPS energies 
($\sqrt{s}_{NN}\approx 20$ GeV), which does not support enough 
multiple scattering among partons to justify a perturbative
treatment. This implies that partonic cascades 
having $p_T > p_T^{\text{min}}$ 
constitute only a small part of the dynamics of the collision at SPS
energies. Indeed we find that the net-baryon multiplicity at central
rapidity in Pb+Pb collision at $\sqrt{s}_{NN} $ = 17.4 GeV is only about
20\% of the experimental value estimated by the NA49 experiment~\cite{na49}.
Similar considerations also apply to $p+p$ collisions, which do not 
produce a dense partonic medium where color screening occurs at a
perturbative scale $p_T^{\text{min}} > \Lambda_{QCD}$. The transport of net
baryon number then must be due to nonperturbative mechanisms. The
situation is dramatically different in $Au+Au$ collisions at RHIC 
energies, where multiple parton scattering at $p_T >p_T^{\text{min}}$ produces 
a medium, in which color is screened at a sufficiently short distance
to allow for a choice of $p_T^{\text{min}}$ in the domain of pQCD.

In conclusion, we find that the parton cascade model predicts a
net baryon excess at mid-rapidity in Au+Au collisions at RHIC,
which is in qualitative agreement with the measured values. 
Two mechanisms are driving this excess: One is the presence of 
a net baryon density in the initial state parton distributions 
at Bjorken-$x$ around 0.01 reflecting the size of the valence 
quark component in this range of $x$. The other important factor
is the rescattering among partons, which transports more partons,
and hence additional net baryon number, to mid-rapidity. This
transport mechanism increases the net baryon number density well 
into the range of measured values. 

\begin{acknowledgments}  
This work was supported in part by RIKEN, the Brookhaven National 
Laboratory, DOE grants DE-FG02-96ER40945 and DE-AC02-98CH10886, 
and by the Natural Sciences and Engineering Research Council of 
Canada. 
\end{acknowledgments}

\end{document}